# Multiple Sources of the European Neolithic: Mathematical Modelling Constrained by Radiocarbon Dates

**K. Davison**[a, *], **P. M. Dolukhanov**[b], **G. R. Sarson**[a], **A. Shukurov**[a] **& G. I. Zaitseva**[c]

[a] *School of Mathematics and Statistics, University of Newcastle upon Tyne, NE1 7RU, U.K.*
[b] *School of Historical Studies, University of Newcastle upon Tyne, NE1 7RU, U.K.*
[c] *Institute for the History of Material Culture, Russian Academy of Sciences, 18 Dvortsovaya Nab., St. Petersburg, 191186, Russia*

## *Abstract*

We present a mathematical model, based on the compilation and statistical processing of radiocarbon dates, of the transition from the Mesolithic to the Neolithic, from about 7,000 to 4,000 BC in Europe. The arrival of the Neolithic is traditionally associated with the establishment of farming-based economies; yet in considerable areas of north-eastern Europe it is linked with the beginning of pottery-making in the context of foraging-type communities. Archaeological evidence, radiocarbon dates and genetic markers are consistent with the spread of farming from a source in the Near East. However, farming was less important in the East; the Eastern and Western Neolithic have distinct signatures. We use a population dynamics model to suggest that this distinction can be attributed to the presence of two waves of advance, one from the Near East, and another through Eastern Europe. Thus, we provide a quantitative framework in which a unified interpretation of the Western and Eastern Neolithic can be developed.

## *1. Introduction*

The transition to the Neolithic was a crucial period in the development of Eurasian societies, defining to a large extent their subsequent evolution. The introduction of agro-pastoral farming, which originated in the Near East about 12,000 years ago and then spread throughout Europe, is usually considered to be a key feature of this transition (Zvelebil, 1996). Yet the Neolithic was not a simple, single-faceted phenomenon. In his early definition of the Neolithic, Sir John Lubbock (1865) specified its main characteristics to be the growing of crops, the taming of animals, the use of polished stone and bone tools, and pottery-making.

Ceramic pottery is one of the defining characteristics of the Neolithic. It is true that there are examples of early farming communities apparently not involved in pottery-making. For example, aceramic Neolithic cultures have been identified in the Levant, Upper Mesopotamia, Anatolia (9800–7500 BC) and also in the Peloponnese (7000–6500 BC) and Thessaly Plain (7300–6300 BC). (All BC dates supplied are radiocarbon dates calibrated using OxCal v3.10 (Bronk Ramsey, 2001) with calibration curve `intcal04.14c`.) Wheat, barley and legumes were cultivated at those sites; permanent houses with stone foundations were used. There is no widespread evidence of pottery (Perles, 2001) but recent excavations have revealed the occurrence of pottery in Thessaly (J.K. Kozlowski, personal communication 27/03/2007). In contrast, the Neolithic in North-Eastern boreal Europe is identified with a sedentary (or seasonally sedentary) settlement pattern, social hierarchy and sophisticated symbolic expression, the use of polished stone and bone tools, large-scale manufacture of ceramic ware, but not with agriculture (Oshibkina, 1996): the subsistence apparently remained based on foraging. This combination of attributes is characteristic of the 'boreal Neolithic'; of these, pottery is in practice the most easily identifiable.

---

[*] Corresponding author.
  *Email address:* kate.davison@ncl.ac.uk
  *Tel No.* +44 (0) 191 2227218

In the present paper we attempt to develop a unified framework describing the spread of both the 'agro-pastoral' and 'boreal' Neolithic. Our quantitative model of the Neolithization is based on the large amount of relevant radiocarbon dates now available.

## *2. Selection of Radiocarbon Dates*

The compilation of dates used in this study to model the spread of the Neolithic in Europe is available upon request from the authors or in table S1 of supplementary information; unlike all other similar studies known to us it includes dates from the East of Europe. We used data from Gkiasta et al. (2003), Shennan and Steele (2000), Thissen et al. (2006) for Southern, Central and Western Europe (SCWE) and Dolukhanov et al. (2005), Timofeev et al. (2004) for Eastern Europe (EE). Our selection and treatment of the dates, described in this section, is motivated by our attempt to understand the spread of agriculture and pottery making throughout Europe.

Many archaeological sites considered have long series of radiocarbon dates: often with 3–10 dates, and occasionally with 30–50. Associated with each radiocarbon measurement is a laboratory error, which after calibration was converted into a calibration error $\sigma_i$. The laboratory error characterises the accuracy of the measurement of the sample radioactivity rather than the true age of the archaeological site (Dolukhanov et al., 2005) and, thus, is often unrepresentatively small, suggesting an accuracy of 30 years on occasion. Therefore, we estimated an empirical minimum error of radiocarbon age determination of the archaeological age and then used it when treating sites with multiple dates. A global minimum error of $\sigma_{min} = 160$ years is obtained from well explored, archaeologically homogeneous sites with a large number of tightly clustered dates. Such sites are: (1) Ilipinar, 65 dates, with the standard deviation $\sigma = 168$ years (and mean date 6870 BC); (2) Achilleion, 41 dates, $\sigma = 169$ years (mean 8682 BC); (3) Asikli Höyük, 47 dates, $\sigma = 156$ years (mean 7206 BC). Similar estimates are $\sigma_{min} = 100$ years for LBK sites and $\sigma_{min} = 130$ years for the Serteya site in North-Western Russia (Dolukhanov et al., 2005); the typical errors vary between different regions and periods but we apply $\sigma_{min} = 160$ years to all the data here.

For sites with multiple radiocarbon date determinations, the dates are treated and reduced to two (and, rarely more) dates that are representative of the arrival of multiple Neolithic episodes to that location. For the vast majority of such sites, the radiocarbon dates available can be combined, as discussed below, to just two possible arrival dates. Examples of sites with multiple radiocarbon measurements are Ilipinar and Ivanovskoye-2 where, respectively, 65 and 21 dates have been published. Figs 1a and b indicate that for these sites the series of dates form very different distributions; different strategies are used to process these different types of date series as described below (see Dolukhanov et al., 2005, for detail). If a geographical location hosts only one radiocarbon measurement associated with the early Neolithic, then this is taken to be the most likely date for the arrival of the Neolithic. The uncertainty of this radiocarbon date is taken to be the maximum of the global minimum error discussed above and the calibrated data range first obtained at the 99.7% confidence level and then divided by six (to obtain an analogue of 1σ error). There are numerous such sites in our collection, including Casabianca, Dachstein and Inchtuthil.

If only a few (less than 8) date measurements are available for a site and those dates all agree within the calibration error, we use their weighted mean value and characterise its uncertainty with an error equal to the maximum of each of the calibrated measurement errors $\sigma_i$, the standard deviation of the dates involved $\sigma(t_i)$, $1 \leq i \leq n$, and the global minimum error introduced above:

$$\sigma = \max\{\sigma_i, \sigma(t_i), \sigma_{min}\}, \tag{1}$$

where $n$ is the total number of dates in the cluster. An example of such a site is Bademağaci, where we have 4 dates, all within 60 years of one another; Figure 1c shows the histogram of radiocarbon dates of this site. The typical calibration error of these dates is approximately 30 years, thus Eq. (1) yields $\sigma_{min}$ as an uncertainty estimate.

For a series of dates that cluster in time but do not agree within the calibration error, we use different approaches depending on the number of dates available and their errors. Should the cluster contain less than 8 dates, we take the mean of the dates (as in the previous case), as any more sophisticated statistical technique would be inappropriate for such a small sample; the error is taken as in Eq. (1). An example of such a site is Okranza Bolnica – Stara Zagora with 7 measurements, Fig. 1f shows that the dates are tightly clustered around the mean value.

If however, the date cluster is large (i.e. more than 8 dates, such as Ilipinar, shown in Fig. 1a), the $\chi^2$ statistical test can be used to calculate the most likely date $T$ of a coeval subsample as described in detail by Dolukhanov et al. (2005):

$$T = \frac{\sum_{i=1}^{n} t_i / \tilde{\sigma}_i^2}{\sum_{i=1}^{n} 1 / \tilde{\sigma}_i^2}.$$

where $\tilde{\sigma}_i = \max(\sigma_i, \sigma_{min})$. The coeval subsample is obtained by calculating the statistic $X^2 = \sum_{i=1}^{n} \frac{(t_i - T)^2}{\tilde{\sigma}_i^2}$ and comparing it with $\chi_{n-1}^2$. If $X^2 \leq \chi_{n-1}^2$, the sample is coeval and the date $T$ is the best representative of the sample. If $X^2 > \chi_{n-1}^2$, the sample is not necessarily coeval, and the dates that provide the largest contribution to $X$ are discarded one by one until the criterion for a coeval sample is satisfied. This process is very similar to that implemented in the *R_Combine* function of OxCal (Bronk Ramsey, 2001). However, OxCal's procedure first combines the uncalibrated dates into one single radiocarbon measurement and only then calibrates it. Our approach on the other hand first uses the calibration scheme of OxCal and then combines the resulting calibrated dates to give $T$. Furthermore, our procedure adds the flexibility of identifying and discarding dates with the largest relative deviation from $T$. Within *R_Combine* the minimum error is not used in the calculation of $X^2$ but is rather only incorporated into the final uncertainty estimate. We feel that it is more appropriate to include the minimum uncertainty into the calculation from the outset. As a check, we combined several sets of dates using both OxCal and our procedure, and the results agree within an acceptable margin (where such agreement could be expected). An example of such a test site is Bouqras (35.50ºN, 40.47ºE) where there are 16 date measurements available spread between 6865 ± 62 BC and 7440 ± 37 BC. Our selection method gives $T = 7195 \pm 198$ BC, whereas OxCal's *R_Combine* results in $T = 7139 \pm 143$ BC. When the minimum error is taken into consideration in the application of the $\chi^2$ test, the dates pass the test and can be combined into a single date given above. *R_Combine* in OxCal, however, concludes that the 16 dates do not form a coeval sample as the $\chi_{n-1}^2$ criterion because of the smaller errors employed in the test. As discussed above, the errors used by OxCal only include those arising from the sample's radioactivity measurements and calibration but neglect all other errors. Then our assertion is that it is more appropriate to incorporate this minimum error in the calculation at all stages as implemented in the procedure used here.

A further method to analyse data sets available in OxCal is the calculation of phase boundaries using Bayesian methods (the *Phase/Boundary* function), where the additional (prior) information used is that on the probability distribution of the dates in the set. However, it can be difficult to formulate such an additional hypothesis in a meaningful manner, so that any additional constraints of this nature would only distort the result. Nevertheless, we conducted some exploratory analysis into the effects of estimating the date of 'the first arrival' for a set of dates in this manner. An alternative, traditional approach, would be to use, merely the earliest date in an extended set of dates without any discernible maxima or, otherwise, the age of the earliest maximum. We will return to this in Section 4, but we consider the example of Bouqras here. Using the *Boundary* function of OxCal, the start of the phase for Bouqras is estimated as 7477 ± 60 BC, which is about 300 years (i.e., about 2σ) earlier than the mean date as determined above.

If a site has many radiocarbon determinations that do not cluster around a single date, a histogram of the dates is analyzed. If the data have a wide range and have no discernable peaks (i.e., are approximately uniformly distributed in time), they may suggest prolonged Neolithic activity at the site, and we choose, as many other authors, the oldest date (or one of the oldest, if there are reasons to reject outliers) to identify the first appearance of the Neolithic. Examples of such sites are Mersin and Halula where there are 6 and 9 dates with a range of 550 and 1900 years, respectively, and no significant peaks (see Figs 1d and 1e), here the oldest dates are 6950 and 8800 years BC and the associated errors are 217 and 167 years.

Apart from sites with either no significant peak or only one peak, there are sites whose radiocarbon dates have a multimodal structure which may indicate multiple waves of settlement passing through this location. Ivanovskoye-2 (with 21 dates) is a typical site in this category, and Fig. 1b depicts two distinct peaks. In such cases multiple dates were attributed to the site, with the above methods applied to each peak independently. Admittedly our method of assigning an individual date to a specific peak could be in some cases inaccurate, as appropriate stratigraphic and/or typological data are not invoked in our procedure. In future refinements to this technique we may consider fitting bimodal normal distributions to the data to avoid the rigid assignment of measurements to one peak or another. After selection and processing, the total number of dates in our compilation is 477. In our final selection 30 sites have two dates allocated and 4 sites have three dates, namely Berezovaya (60.38ºN, 44.17ºE), Osipovka (49.93ºN, 30.40ºE), Rakushechnyi Yar (47.55ºN, 40.67ºE) and Yerpin Pudas (63.35ºN, 34.48ºE). A table of all of the data used in this paper is available as supplementary online information.

## 3. Modelling

The mechanisms of the spread of the Neolithic in Europe remain controversial. Gordon Childe (1925) advocated direct migration of the farming population; this idea was developed in the form of the demic expansion (wave of advance) model (Ammerman and Cavalli-Sforza, 1973). The Neolithization was viewed as the spread of colonist farmers who overwhelmed the indigenous hunter-gatherers or converted them to the cultivation of domesticated cereals and the rearing of animal stock (Price, 2000). An alternative approach views the Neolithization as an adoption of agriculture (or other attributes) by indigenous hunter-gatherers through the diffusion of cultural novelties by means of intermarriages, assimilation and borrowing (Thomas, 1996; Tilley, 1994; Whittle, 1996). Recent genetic evidence seems to favour cultural transmission (Haak et al., 2005).

Irrespective of the particular mechanism of the spread of the Neolithic (or of its various signatures), the underlying process can be considered as some sort of 'random walk', of either humans or ideas and technologies. Therefore, mathematical modelling of the spread (at suitably large scales in space and time) can arguably be based on a 'universal' equation (known as reaction-diffusion equation) with parameters chosen appropriately (Cavalli-Sforza and Feldman, 1981). A salient feature of this equation is the development of a propagation front (where the population density, or any other relevant variable, is equal to a given constant value) which advances at a constant speed (Murray, 1993) (in the approximation of a homogeneous, one-dimensional habitat). This mode of spread of incipient agriculture has been confirmed by radiocarbon dates (Ammerman and Biagi, 2003; Ammerman and Cavalli-Sforza, 1971; 1973; 1984; Gkiasta et al., 2003; Pinhasi et al., 2005). In Fig. 2a we plot the distance from a putative source in the Near East versus the $^{14}$C dates for early Neolithic sites in SCWE; the linear interdependence is consistent with a constant propagation speed. Due to the inhomogeneous nature of the landscape we would not expect to see a very tight correlation between distance from source and time of first arrival, since there are many geographical features that naturally cause barriers to travel (e.g. the Mediterranean Sea). It is also suggested in a previous work (Davison et al., 2006) that there are local variations in the propagation speed near major waterways, this again detracts from the constant rate of spread. In spite of this the correlation coefficient is found to be -0.80, reassuringly high given the above complications. There is also a tail of older dates that originate in early Neolithic sites in the Near East, where a Neolithic tradition began and remained until it saturated the area and subsequently expanded across the landscape.

In contrast to earlier models, we include the 'boreal', East-European (EE) Neolithic sites, which we present in the same format in Fig. 2b. It is clear that the Eastern data are not all consistent with the idea of spread from a single source in the Near East. A correlation coefficient of -0.52 between the EE dates and distance to the Near East is sufficient evidence for that. Our modeling, discussed below, indicates that another wave of advance swept westward through Eastern Europe about 1500 years earlier than the conventional Near-Eastern one; we speculate that it may even have spread further to produce early ceramic sites in Western Europe (e.g. the La Hoguette and Roucadour groups).

Our population dynamics model, described in detail by Davison et al. (2006), was refined for our present simulations. The model is based on the random walk of individuals first considered in a similar context by Fisher (1937). At any point in space each individual will take a step in any given direction with the same probability, i.e. they are as likely to step left as they are right, forwards or backwards. This assumption of equal probabilities gives rise to an isotropic random walk, i.e. classical diffusion. If however the probability of moving in one direction is altered by a desire for a particular environment or the ease of travel in a certain direction the random walk becomes anisotropic. We thus solve the reaction–diffusion equation supplemented with an advection of speed **V**, arising from this anisotropic component of the random walk of individuals that underlies the large-scale diffusion (Davison et al., 2006; Murray, 1993):

$$\frac{\partial N}{\partial t} + (\mathbf{V} \cdot \nabla) N = \gamma N \left(1 - \frac{N}{K}\right) + \nabla \cdot (\nu \nabla N), \tag{2}$$

where $N$ is the population density, $\gamma$ is the intrinsic growth rate of the population, $K$ is the carrying capacity, and $\nu$ is the diffusivity (mobility) of the population. We solve Eq. (2) numerically in two dimensions on a spherical surface with grid spacing of 1/12 degree (2–8 km, depending on latitude). All the variables in Eq. (2) can be functions of position and time, as described below and by Davison et al. (2006).

We consider two non-interacting populations, each modelled with Eq. (2), but with different values of the parameters **V**, $\gamma$, $K$ and $\nu$; the difference is intended to represent differences between subsistence strategies (farmers versus hunter-gatherers) and/or between demic and cultural diffusion.

We thus numerically solve two versions of Equation (2), one for each of two populations with different origins of dispersal. The numerical scheme adopted has centered differences in space and evolves with explicit Euler time stepping using forward differences in time. The size of the time step $\Delta t$ is controlled using the Courant, Friedrichs, Lewy (CFL) condition, where the population front is prevented from advancing more than one grid cell in one time step:

$$\Delta t \leq \min\left\{\frac{A_1}{2\nu} R^2 \frac{\Delta \phi^2 \Delta \theta^2}{\Delta \phi^2 + \Delta \theta^2}, \frac{A_2}{2|V_\phi|} R \Delta \phi, \frac{A_3}{2|V_\theta|} R \Delta \theta\right\}, \quad 0 < A_i \leq 1,$$

where $\Delta \phi$ and $\Delta \theta$ are the mesh sizes in the azimuth and latitude, $R$ is the Earth radius, $V_\phi$ and $V_\theta$ are the azimuthal and latitudinal component of the advection velocity **V**, respectively. Here we cautiously take $A_i$=0.01. The boundaries of the computational domain are at 75ºN and 25ºN, and 60ºE and 15ºW as shown in Fig. 4, they are chosen to comfortably incorporate our pan-European area. We use zero-flux boundary conditions, i.e. $\partial N / \partial \mathbf{n} = 0$, where **n** is the normal to the boundary (although in most cases the boundary condition hardly affects the result as the boundary is in the sea). The environmental factors included into the model are the altitude, latitude, coastlines and the Danube-Rhine river system. The equation describing the farming population also includes advection velocity **V** along the major waterways (the Danube, the Rhine and the sea coastlines; **V** ≠ 0 within corridors 10 km wide on each side of a river or 10 km inshore near the sea) which results from anisotropic diffusion in those areas. Detailed prescription of the components of the advective velocity is given in Davison et al. (2006); briefly, |**V**| diminishes with distance from coastlines and major rivers; the direction of the velocity is taken to be along the river or coastline and away from the maximum of the population. The magnitude of **V** is considered below.

The focus of our model is the speed of the front propagation *U,* since this quantity can be most readily linked to the radiocarbon age used to date the 'first arrival' of the wave of advance. This feature of the solution depends only on the linear terms in Equation (2) and, in particular, is independent of the carrying capacity *K*. Moreover, to a first approximation *U* only depends on the product γν:

$$U = 2\sqrt{\gamma \nu} \ . \tag{3}$$

Taking the intrinsic growth rate of a farming population as γ = 0.02 year$^{-1}$ (Birdsell, 1957), the mean speed of the front propagation of $U \approx 1$ km/year for the population of farmers suggests the background (low-latitude) value of the diffusivity ν = 12.5 km²/year (Ammerman and Cavalli-Sforza, 1971; Davison et al., 2006). For the wave spreading from Eastern Europe, $U \approx 1.6$ km/year is acceptable as a rough estimate obtained from the EE radiocarbon dates (Dolukhanov et al., 2005); this estimate is confirmed by our model (see Fig. 2d). Previous analysis of the spread of Paleolithic hunter-gatherers yields $U \approx 0.8$ km/year; the corresponding demographic parameters are suggested to be γ = 0.02–0.03 year$^{-1}$ and ν = 50–140 km²/year (Fort et al., 2004). These authors use an expression for *U* different from Eq. (3); it is plausible, therefore, that the intrinsic growth rate obtained by Fort et al. (2004) for hunter-gatherers is a significant overestimate; for ν = 100 km²/year and $U \approx 1.6$ km/year, the nominal value of γ obtained from Eq. (3) is about 0.006 year$^{-1}$. A growth rate of γ = 0.01 year$^{-1}$ has been suggested for indigenous North-American populations in historical times (Young and Bettinger, 1992). The range γ = 0.003–0.03 year$^{-1}$ is considered in a model of Paleoindian dispersal (Steele et al., 1998). Our simulations adopt γ = 0.007 year$^{-1}$ and ν = 91.4 km²/year for the hunter-gatherers.

For the wave that spreads from the Near East carrying farming, *K* and ν smoothly tend to zero within 100 m of the altitude 1 km, above which land farming becomes impractical. For the wave spreading from the East, *K* and ν are similarly truncated at altitudes around 1,500 km as foraging is possible up to higher altitude than farming (see Fig. 3b). The low-altitude (background) values of *K* adopted are 0.07 persons/km² for hunter-gatherers (Dolukhanov, 1979; Steele et al., 1998) and 3.5 persons/km² for farmers, a value 50 times larger than that for hunter-gatherers (Ammerman and Cavalli-Sforza, 1984). The values of *K* do not affect any results reported in this paper.

In seas, for both farmers and hunter-gatherers, both the intrinsic growth rate and the carrying capacity vanish as seas are incapable of supporting a human population. The diffusivity for both farmers and hunter gatherers tails off exponentially as $\nu \propto \exp(-d/l)$, with *d* the shortest distance from the coast and *l* = 40 km, allowing the population to travel within a short distance off shore but not to have a sustained existence there. The value of *l* has been fine-tuned in this work in order to reproduce the delay, indicated by radiocarbon dates, in the spread of the Neolithic from the continent to Britain and Scandinavia (see Fig. 3a). This provides an interesting inference regarding the sea-faring capabilities of the times, suggesting confident travel within about 40 km off the coast. We have used the present-day sea levels although we acknowledge that the sea level has changed over the past 6,000 years. However, the variation in the position of the shoreline is effectively allowed for by the fact that the diffusivity in our model decreases only gradually off the modern coastline.

The inclusion of advection along the Danube–Rhine corridor and the sea coastlines is required to reproduce the spread of the Linear Pottery and Impressed Ware cultures obtained from the radiocarbon and archaeological evidence (see Davison et al., 2006, for details). The speed of spread of farming in the Danube–Rhine corridor was as high as 4 km/yr (Ammerman and Cavalli-Sforza, 1971) and that in the Mediterranean coastal areas was perhaps as high as 20 km/yr (Zilhão, 2001); we set our advective velocity in these regions accordingly. However, there are no indications that similar acceleration could occur for the hunter-gatherers spreading from the East. Thus, we adopt **V** = 0 for this population.

The starting positions and times for the two waves of advance — i.e., the initial conditions for Eq. (2) — were selected as follows. For the population of farmers, we position the origin and adjust the starting time so as to minimize the root mean square difference between the SCWE $^{14}$C dates and the arrival time of the modelled population at the corresponding locations; the procedure is repeated for all positions between 30ºN, 30ºE and 40ºN, 40ºE, with a 1º step. This places the centre at 35ºN, 39ºE, with the propagation starting at 6,700 BC. For the source in the East of Europe, we have tentatively selected a region centered at 53ºN, 56ºE in the Ural mountains (to the east of the Neolithic sites used here), so that the propagation front reaches the sites in a well developed form. We do *not* suggest that pottery-making independently originated in this region. More reasonably, this technology spread, through the bottleneck between the Ural Mountains and the Caspian Sea, from a location further to the east. The starting time for this wave of advance was fixed by trial and error at 8200 BC at the above location; this reasonably fits most of the dates in Eastern Europe attributable to this centre. For both populations, the initial distribution of *N* is a truncated Gaussian of a radius 300 km.

## *4. Comparison of the model with radiocarbon dates*

The quality of the model was assessed by considering the time lag $\Delta T = T - T_m$ between the modelled arrival time(s) of the wave(s) of advance to a site, $T_m$, and the actual $^{14}$C date(s) of this site, *T*, obtained as described in Sect. 2. The sites were attributed to that centre (Near East or Urals) which provided the smallest magnitude of Δ*T*. This procedure admittedly favours the model, and the attributions have to be carefully compared with the archaeological and typological characteristics of each site. Such evidence is incomplete or insufficient in a great number of cases; we leave the laborious task of incorporating independent evidence in a systematic and detailed manner for future work. Our formulaic method of attribution has inevitably failed in some cases, but our preliminary checks have confirmed that the results are still broadly consistent with the evidence available (see below).

First, we considered a model with a single source in the Near East. The resulting time lags are presented in Fig. 4a–c. The best-fit model with two sources is similarly illustrated in Fig. 4d-f. The locations of the sources are shown with grey ellipses in panels (c) and (f).

In Fig. 4a the sites shown are those at which the model arrival date and the radiocarbon date agree within 500 years (55% of the pan-European dates); Fig. 4d gives a similar figure for the two source model (now 70% of the pan-European dates fit within 500 years). The points in the EE area are significantly more abundant in Fig. 4d than in Fig. 4a, while the difference in the SCWE area is less striking. The SCWE sites are better fitted with the one source model, with |Δ*T*| < 500 years for 68% of data points, but the fit is unacceptably poor for EE, where only 38% of the radiocarbon dates can be fitted within 500 years. A convenient measure of the quality of the fit is the standard deviation of the time lags

$$s = \sqrt{\frac{1}{N} \sum_{i=1}^{N} \left( \Delta T_i - \overline{\Delta T} \right)} \quad \text{with} \quad \overline{\Delta T} = \frac{1}{N} \sum_{i=1}^{N} \Delta T_i.$$

The standard deviation of the pan-European time lags here is *s* = 800 years. Outliers are numerous when all of the European sites are included (illustrated by the abundance of points in Fig. 4c), and they make the distribution skewed, and offset from Δ*T* = 0. The outliers are mainly located in the east: for the SCWE sites, the distribution is more tightly clustered (*s* = 540 years), has negligible mean value, and is quite symmetric. In contrast, the time lags for sites in Eastern Europe (EE), with respect to the centre in the Near East, have a rather flat distribution (*s* = 1040 years), which is strongly skewed and has a significant mean value (310 years). The failure of the single-source model to accommodate the $^{14}$C dates from Eastern Europe justifies our use of a more complicated model that has two sources of propagation. Attempts were made at locating the single source in various other locations, such as the Urals, but this did not improve the agreement.

Adding another source in the East makes the model much more successful: the values of the time lag, shown in Fig. 4d–f, are systematically smaller; i.e. there are significantly fewer points in Fig. 4f (5%) compared to Fig. 4c (17%). The resulting distribution of Δ*T* for all the sites is quite narrow ($s = 520$ years) and almost perfectly symmetric, with a negligible mean value (40 years). The distributions remain similarly acceptable when calculated separately for each source (with $s = 490$ and 570 years for the sites attributable to the Near East and Urals, respectively). The improvement is especially striking in EE, where the sites are split almost equally between the two sources.

We tentatively consider a model acceptable if the standard deviation, $s$, of the time lag $\Delta T$ is not larger than 3 standard dating errors σ, i.e., about 500 years, given our estimate of σ close to 160 years over the pan-European domain. This criterion cannot be satisfied with any single-source model, but is satisfied with two sources. While we would never expect a large-scale model of the sort proposed here to accurately describe the complex process of the Neolithization in fine detail (and so the resulting values of $\Delta T$ cannot be uniformly small), the degree of improvement in terms of the standard deviation of $\Delta T$ clearly favours the two-source model. The reduction in $s$ is statistically significant, and cannot be explained by the increase in the complexity of the model alone. The confidence intervals of the sample standard deviations $s$ for one-source and two-source models do not overlap ($740 < \sigma < 840$ and $480 < \sigma < 550$, respectively); the F-test confirms the statistical significance of the reduction at a 99% significance level.

It is instructive to examine those dates where our model most strongly deviates from the radiocarbon data, and consider whether our treatment of the radiocarbon dates may be refined (for example by using the *Boundary* function of OxCal or otherwise). Table 1 shows the twelve most strongly deviating dates, i.e., those with the largest value of $(\Delta T)^2 / \sigma^2$. Of those twelve cases, our model arrives too late on seven occasions. Of these seven sites only one has not undergone statistical treatment, the other six are all sites at which the process described in Section 2 led us to select the oldest date at a site. This is, of course, the least reliable method to estimate the date of the 'first arrival' from a group of dates. These sites may thus require further examination. The five sites in Table 1 where our model arrives too early are those at which only one radiocarbon measurement was available, so that no statistical treatment has been applied. It can also be noted quantitatively from Table 1 and qualitatively from Fig. 4f that the sites where our model most strongly deviates from the radiocarbon date are uniformly distributed on the map. Thus, no regional bias is evident in the fit, which suggests that any model refinement via manipulation of the parameters would not increase its accuracy. An improvement in the accuracy of modelling would require the use of more detailed, regional models based on, say, stochastic equations or direct modelling of the random-walk. Incidentally, the use of the *Boundary* function of OxCal in order to estimate the 'first arrival' time would make the deviation of the model from the data stronger for the seven sites where the modelled arrival time is younger than the available radiocarbon date.

It is instructive to represent the data in the same format as in Figs 2a,b, but now with each date attributed to one of the sources, as suggested by our model. This has been done in Figs 2c, d, where the close correlation of Fig. 2a is restored for the pan-European data. Now, the dates are consistent with constant rates of spread from one of the two sources. Using straight-line fitting, we obtain the average speed of the front propagation of 1.1 ± 0.1 km/year for the wave originating in the Near East (Fig. 2c), and 1.7 ± 0.3 km/year for the source in the East (Fig. 2d); 2σ values are given as uncertainties here and below. The spread from the Near East slowed down in Eastern Europe to 0.7 ± 0.1 km/year; excluding the dates from the west alone (as in Fig. 2a) gives a higher speed of 1.2 ± 0.1 km/year. The estimates for the data in both western and eastern Europe are compatible with earlier results (Dolukhanov et al., 2005; Gkiasta et al., 2003; Pinhasi et al., 2005). Care must be taken when using such estimates, however, since the spread occurs in a strongly heterogeneous space, and so cannot be fully characterised by a single constant speed. The rate of spread varies on both pan-European scale and on smaller scales, e.g., near major waterways (Davison et al., 2006).

Our allocation of sites to the sources of initial spread discussed above clearly needs careful verification using independent evidence. Here we briefly discuss a few sites. Taking Ivanovskoye-2 (56.85ºN, 39.03ºE) as an example, the data form two peaks shown in Fig. 1b; the times at which each of the waves arrives at this location are 4349 BC for the Near-Eastern wave and 5400 BC for the Eastern wave, closely fitting the two peaks in $^{14}$C dates. As another example, we accept two dates for Mayak (68.45ºN, 38.37ºE); one from the younger cluster (2601 ± 192 BC), and also the older date (4590 ± 47 BC) detached from the cluster. The younger cluster is consistent with the Near-Eastern wave (arriving at 2510 BC) and the older date agrees with the Eastern wave arriving at 4720 BC.

There are also some sites with multiple-peaked histograms of radiocarbon dates, which do not fit our scheme of allocation. As an example, Koshinskaya (27.63ºN, 48.23ºE) has two peaks giving dates of 3550 ± 167 BC and 7360 ± 73. However the Near-Eastern wave arrives at 3900 BC whereas the Eastern wave reaches the site at 5767 BC. While the agreement with the wave arriving from the Near East is acceptable, the older radiocarbon date is too old to fit the other wave. In such cases, we feel that the relevance of the older radiocarbon date has to be carefully reconsidered. There are also sites such as Serteya with dates 3688 ± 200 BC and 6225 ± 317 BC, but the Near-Eastern wave arrives at 4571 BC, and Eastern one, at 5081 BC. Here both dates do not fit the wave.

## *5. Conclusions*

Our model has significant implications for the understanding of the Neolithization of Europe. It substantiates our suggestion that the spread of the Neolithic involved at least two waves propagating from distinct centres, starting at about 8200 BC in Eastern Europe and 6700 BC in the Near East. The earlier wave, spreading from the east via the 'steppe corridor', resulted in the establishment of the 'eastern version' of the Neolithic in Europe. A later wave, originating in the Fertile Crescent of the Near East, is the better-studied process that brought farming to Europe.

It is conceivable that the westernmost extension of the earlier (eastern) wave of advance produced the pre-agricultural ceramic sites of La Hoguette type in north-eastern France and western Germany, and Roucadour-type (also known as Epicardial) sites in western Mediterranean and Atlantic France (Berg and Hauzer, 2001; Jeunesse, 1987). The available dates for the earlier Roucadour sites (7500–6500 BC) (Roussault-Laroque, 1990) are not inconsistent with this idea, but a definitive conclusion needs additional work. Examples of sites in the west allocated to the Eastern source are: Bridgemere (51.21ºN, 2.41ºW), Cherhill (51.43ºN, 1.95ºW), Feldbach (47.23ºN, 8.78ºE), Frankenau (47.50ºN, 16.50ºE), Phyn (47.58ºN, 8.93ºE) and Zurich (47.37ºN, 8.58ºE). This attribution of our model has to be carefully verified using archaeological information.

The nature of the Eastern source needs to be further explored. The early-pottery sites of the Yelshanian Culture (Mamonov, 2000) have been identified in a vast steppe area stretching between the Lower Volga and the Ural Rivers. The oldest dates from that area are about 8000 BC (although the peak of the culture occurred 1000 years later) (Dolukhanov et al., 2005). Even earlier dates have been obtained for pottery bearing sites in Southern Siberia and the Russian Far East (Kuzmin and Orlova, 2000; Timofeev et al., 2004). This empirical relation between our virtual Eastern source and the earlier pottery-bearing sites further east may indicate some causal relationship.

According to our model, the early Neolithic sites in Eastern Europe belong to both waves, in roughly equal numbers. Unlike elsewhere in Europe, the wave attributable to the Near East does not seem to have introduced farming in the East. The reason for this is not clear and may involve the local environment where low fertility of soils and prolonged winters are combined with the richness of aquatic and terrestrial wildlife resources (Dolukhanov, 1996).

Regardless of the precise nature of the eastern source, the current work suggests the existence of a wave which spread into Europe from the east carrying the tradition of early Neolithic pottery-making. If confirmed by further evidence (in particular, archaeological, typological, and genetic), this suggestion will require serious re-evaluation of the origins of the Neolithic in Europe.

## *Acknowledgements*

Financial support from the European Community's Sixth Framework Programme under the grant NEST-028192-FEPRE is acknowledged.

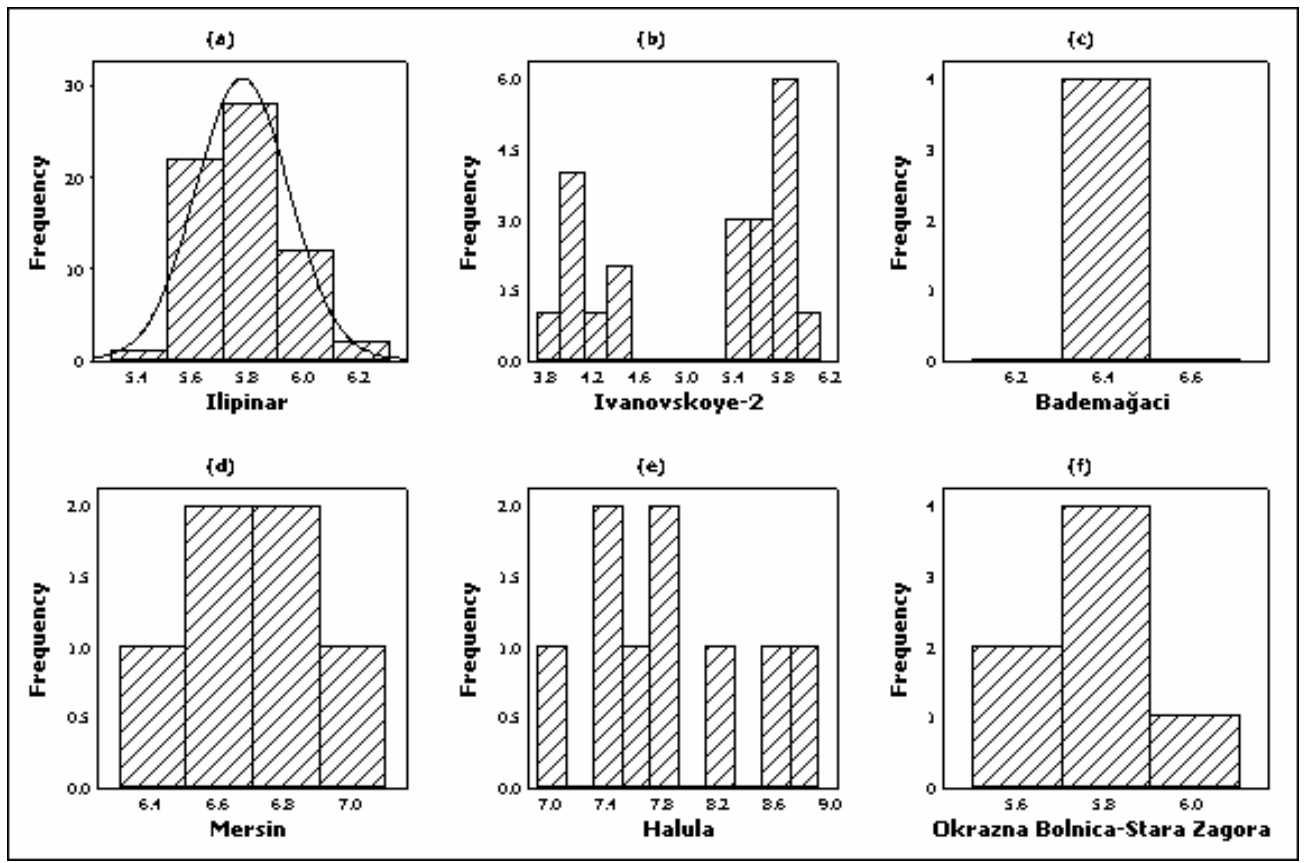

**Figure 1** Histograms of calibrated radiocarbon ages from archaeological sites in kyr BC, binned into 200 year intervals representing various temporal distributions. (a) The 65 dates from Ilipinar (40.47ºN, 29.30ºE) are approximately normally distributed, so the $\chi^2$ criterion can be employed to calculate the age of this site as described by Dolukhanov et al (2005). The resulting Gaussian envelope is shown solid. (b) Ivanovskoye-2 (56.85ºN, 39.03ºE) has 21 dates showing a multimodal structure where each peak can be treated as above. (c) The 4 dates from Bademağaci (37.40ºN, 30.48ºE) combine into a single date when their errors are taken into account. (d) The 6 dates from Mersin (36.78ºN, 34.60ºE) are almost uniformly distributed in time, so the oldest date can be used as representative of the arrival of the Neolithic. (e) The 9 dates from Halula (36.40ºN, 38.17ºE) are treated as in (d). (f) The 7 dates from Okrazna Bolnica – Stara Zagora (42.43ºN, 25.63ºE) are not numerous enough to justify the application of the $\chi^2$ test, but they form a tight cluster, so the mean date can be used for this site.

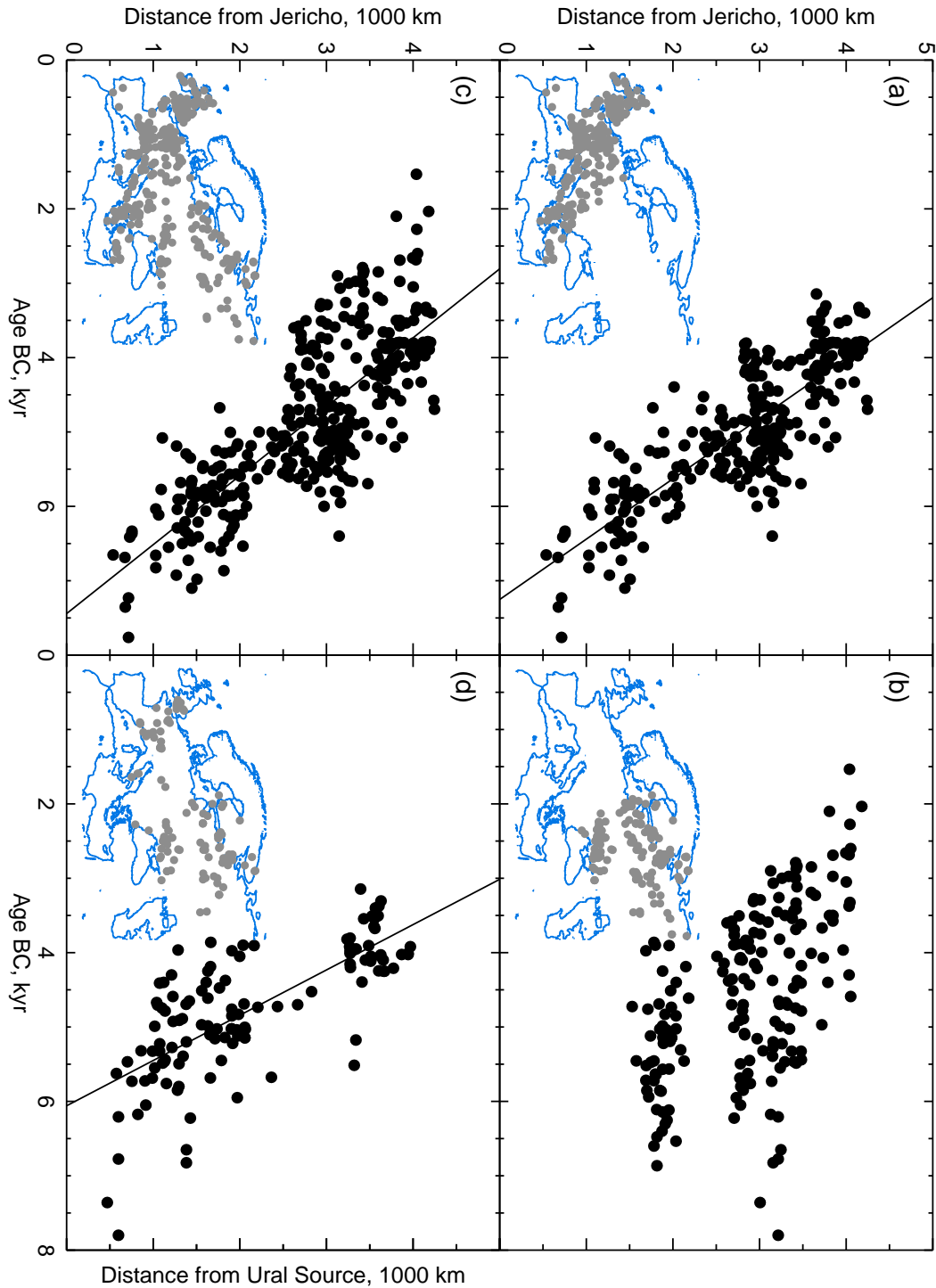

**Figure 2** Radiocarbon dates of early Neolithic sites versus the great-circle distance from the assumed source. Inset maps show the location of the sites plotted, and the straight lines correspond to spread at a constant speed given below. **(a)** Sites from Southern, Central and Western Europe (SCWE) with respect to a Near Eastern source (Jericho). The linear correlation (cross-correlation coefficient $C = -0.80$) suggests a mean speed of advance of $U = 1.2 \pm 0.1$ km/year ($2\sigma$ error). **(b)** Sites from Eastern Europe (EE) show very poor correlation with respect to the same Near-Eastern source ($C = -0.52$), so that straight-line fitting is not useful. **(c)** Sites attributed, using our two-source model, to the Near-Eastern source (note a significant number of EE sites clearly visible in the inset map) show a reasonable correlation ($C = -0.77$) and a mean speed $U = 1.1 \pm 0.1$ km/year. **(d)** Sites attributed to the Eastern source (from both EE and SCWE) show a correlation similar to that of Panel (c) ($C = -0.76$), and a mean speed $U = 1.7 \pm 0.3$ km/year.

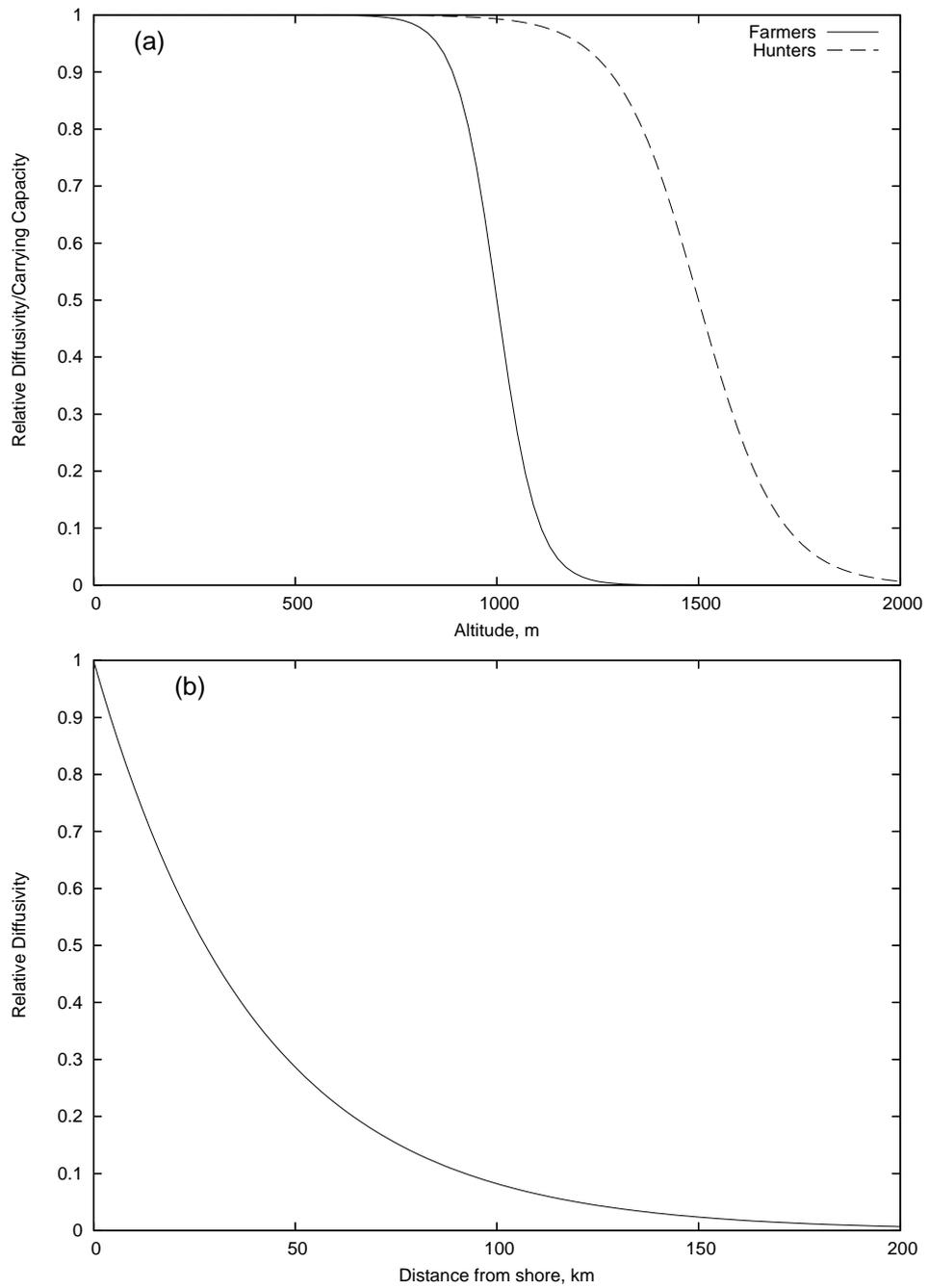

**Figure 3** (a) The dependence of diffusivity and carrying capacity on altitude. (b) The dependence of diffusivity on distance from seashore.

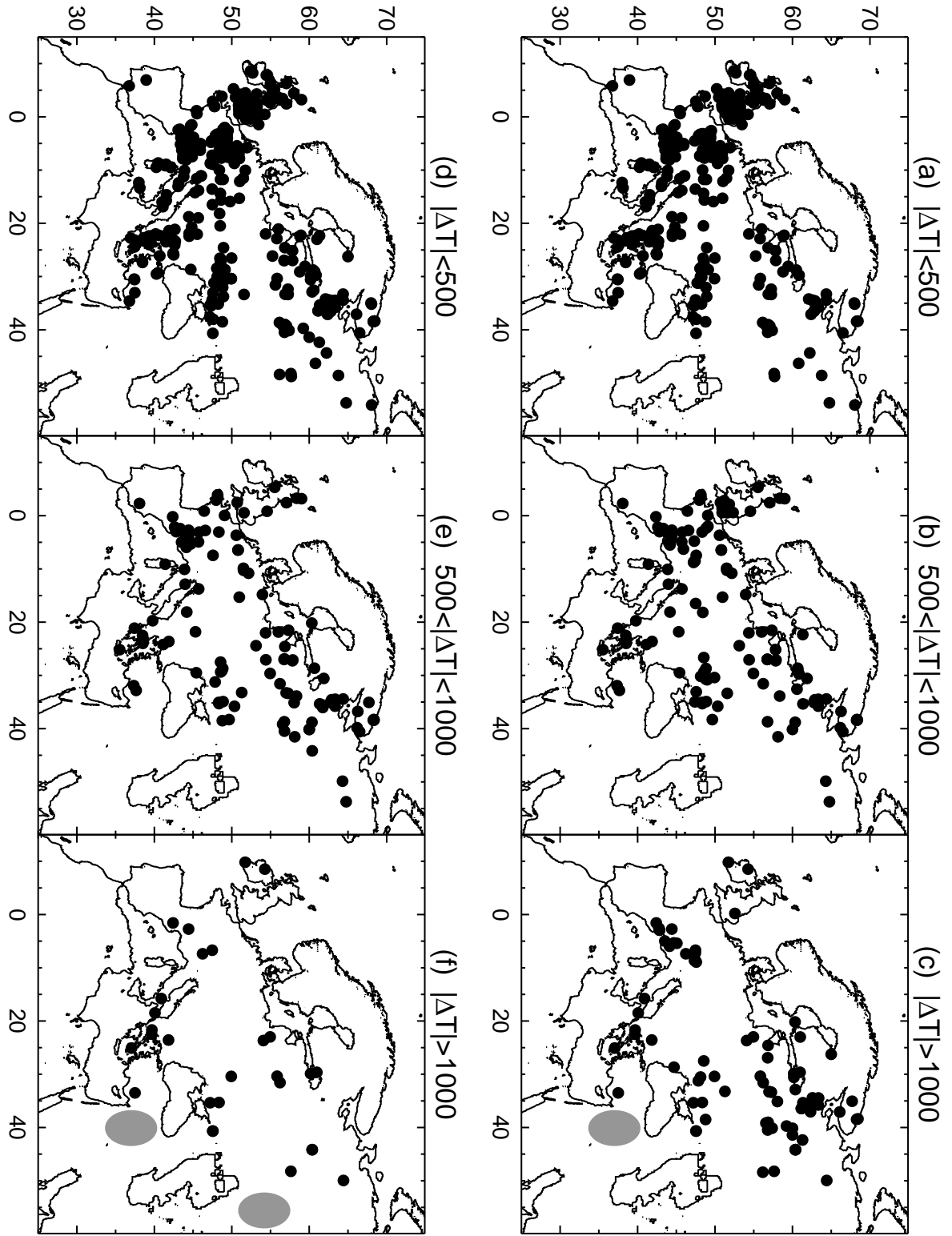

**Figure 4** Time lags, $\Delta T = T - T_m$, between the actual and modelled arrival times for the early Neolithic sites shown against their geographical position: panels **(a)**–**(c)** refer to a model with a single source in the Near East, and panels **(d)**–**(f)** to our best model with two sources (with the second on the Eastern edge of Europe). The positions of the sources are shown in grey in panels (c) and (f). Sites with $|\Delta T| < 500$ are shown in (a) and (d), those with 500 yr < $|\Delta T| < 1000$ yr in panels (b) and (e), and those with $|\Delta T| > 1000$ yr in panels (c) and (f). There are 265, 132, 81 sites in panels (a)–(c) and 336, 116, 26 sites in (d)–(f), respectively. Many data points corresponding to nearby sites overlap, diminishing the apparent difference between the two models. The advantage of the two-source model is nevertheless clear and significant.

| | | | | Sample | | | | | | Model arrival time | |
|---|---|---|---|---|---|---|---|---|---|---|---|
| Site Name | Lab Index | Latitude degrees | Longitude degrees | Age BP, yr | Error | Age cal BC, yr | Calibration Error | Material | Note | From Near East, yr BC | From Urals, yr BC |
| Balma Margineda | Ly-2439 | 42 | 2 | 6670 | 85 | 5600 | 130 | Not given | One date | 6700 | 3256 |
| Berezovaya | LE-67066 | 60.38 | 44.17 | 8700 | 300 | 7800 | 267 | Birch bark | Oldest date | 3697 | 5509 |
| Berezovaya | LE-6706a | 60.38 | 44.17 | 7840 | 75 | 6775 | 92 | Charcoal | Older date | 3697 | 5509 |
| Bolshoe Zavetnoye | LE-6556 | 60.98 | 29.63 | 7750 | 180 | 6650 | 150 | Charcoal | Oldest date | 3836 | 4913 |
| Dobrinišče | Bln-3785 | 41.83 | 23.57 | 6650 | 60 | 5575 | 32 | charcoal | One date | 6700 | 4199 |
| Golubjai-1 | LE-4714 | 54.95 | 22.98 | 7060 | 270 | 5950 | 167 | Charcoal | Older date | 4577 | 4703 |
| Grotta del Sant della Madonna | R-284 | 41 | 16 | 5555 | 75 | 4395 | 155 | Not given | One date | 5722 | 3326 |
| Grotta di Porto Badisco | R-1225 | 40.08 | 18.48 | 5850 | 55 | 4675 | 135 | Not given | One date | 5815 | 3452 |
| Koshinskaya | LE-6629 | 57.63 | 48.23 | 8350 | 100 | 7360 | 73 | Charcoal | Older date | 3896 | 5767 |
| Kurkijokki | LE-6929 | 60.18 | 29.88 | 7900 | 80 | 6825 | 75 | Charcoal | One date | 3973 | 4963 |
| Marevka | OxA-6199 | 48.35 | 35.30 | 7955 | 55 | 6865 | 62 | Bone | Older date | 5566 | 5042 |
| Planta | CRG-280 | 46.23 | 7.37 | 6500 | 80 | 5465 | 155 | Not given | One date | 6700 | 3697 |

**Table 1**  Sites whose 14C dates show the strongest deviation from the model: (1) site name; (2) laboratory index; geographical (3) latitude and (4) longitude in degrees; (5) uncalibrated age and (6) it's 1σ laboratory error in years (BP); (7) calibrated age and (8) it's 1 σ error in years (BC); (9) the sample material; (10) method used to select this date; and the model arrival times (years BC) for the wave spreading from (11) the Near East and (12) the Urals. The data are presented in alphabetical site name order.